\documentclass[seceq]{ptptex}

\usepackage{graphicx}

\title{
Relativistic Hydrodynamics at RHIC and LHC%
}

\author{
Tetsufumi \textsc{Hirano}$^{1,}$\footnote{ e-mail address:
hirano@phys.s.u-tokyo.ac.jp}  
}

\inst{
$^1$ Department of Physics, The University of Tokyo,
Tokyo 113--0033, Japan\\
}

\abst{
Recent development of a hydrodynamic model
is discussed by putting an emphasis on realistic treatment of the early and late stages
in relativistic heavy ion collisions.
The model, which incorporates a hydrodynamic description
of the quark-gluon plasma with a kinetic approach of hadron cascades,
is applied to analysis of elliptic flow data at the Relativistic Heavy Ion Collider energy.
It is predicted that the elliptic flow parameter
based on the hybrid model increases
with the collision energy
up to the Large Hadron Collider energy.
}

\begin{document}

\maketitle

\section{Introduction}

One of the important discoveries made at the Relativistic Heavy Ion  
Collider (RHIC) in Brookhaven National Laboratory
is that the elliptic flow parameter,\cite{Ollitrault:1992bk} namely,
the second Fourier coefficient $v_2 = \langle \cos(2\phi) \rangle$
of the azimuthal momentum distribution $dN/d\phi$,\cite{PoskanzerVoloshin}
is quite large in non-central Au+Au collisions.\cite{experiments}
Over the past years, many studies have been devoted to understanding
the elliptic flow from dynamical models:
(1) The observed
$v_2$ values near midrapidity at low transverse momentum ($p_T$) in 
central and semi-central collisions are consistent with predictions from ideal 
hydrodynamics \cite{Kolb:2000fh}.
(2) The $v_2$ data
cannot be interpreted by hadronic cascade models.\cite{hadroncascade,isse}
(3) A partonic cascade model \cite{MPC} can reproduce
these data only with
significantly larger cross sections than the ones obtained from
the perturbative calculation of quantum chromodynamics.
The produced dynamical system is beyond the
description of naive kinetic theories.
Thus, a paradigm of the strongly coupled/interacting/correlated
matter is being established in the physics of relativistic heavy ion collisions. \cite{sQGP}
The agreement between hydrodynamic predictions and
the data suggests that the heavy ion collision
experiment indeed provides excellent opportunities
for studying matter in local equilibrium at high temperature
and for drawing information of the bulk and transport properties
of the quark-gluon plasma (QGP). 
These kinds of phenomenological studies closely connected with
experimental results, so to say, the ``observational QGP physics",
will be one of the main trends in modern nuclear physics
in the eras of the RHIC and the upcoming Large Hadron Collider (LHC).
Then it is indispensable to sophisticate hydrodynamic modeling
of heavy ion collisions for making quantitative
statements on properties of the produced matter with estimation of uncertainties.
In fact, the ideal fluid dynamical description
gradually breaks down as one studies peripheral collisions \cite{Kolb:2000fh}
or moves away from midrapidity.\cite{Hiranov2eta,HiranoTsuda}
This requires a more realistic treatment 
of the early and late stages \cite{BassDumitru,Teaney,HHKLN,NonakaBass}
in dynamical modeling of relativistic 
heavy ion collisions.

In this paper, recent studies of the state-of-the-art
hydrodynamic simulations are highlighted with emphases on the importance of
the final decoupling stage (Sec.~\ref{sec:success}) and 
of much better understanding of initial conditions (Sec.~\ref{sec:challenge}).
A prediction of the $v_2$ parameter at the LHC
energy will be made in Sec.~\ref{sec:lhc}.
See also other reviews \cite{hydroreviews}
to complement other topics of hydrodynamics in heavy ion collisions at RHIC.

\section{A QGP fluid + hadronic cascade model}
\label{sec:model}

We have formulated a dynamical and unified model, \cite{HHKLN} based on 
fully three-dimensional (3D) ideal hydrodynamics, \cite{Hiranov2eta,HiranoTsuda} 
toward understanding the bulk and transport properties of the QGP.
During the fluid dynamical evolution one assumes local 
thermal equilibrium. However, this assumption can be expected to hold 
only during the intermediate stage of the collision. In order to extract 
properties of the QGP from experimental data one must therefore supplement
the hydrodynamic description by appropriate models for the beginning and
end of the collision.
In Sec.~\ref{sec:success}, we employ the Glauber model for initial conditions
in hydrodynamic simulations.
Initial entropy density is parametrized as a superposition of terms 
scaling with the densities of participant nucleons and binary collisions,
suitably generalized to account for the longitudinal structure of the initial 
fireball.\cite{HHKLN}
Instead, in Secs.~\ref{sec:challenge} and \ref{sec:lhc},
we employ the Color Glass 
Condensate (CGC) picture \cite{Iancu:2003xm} for colliding nuclei and calculate the
produced gluon distributions \cite{KLN} as input for the initial conditions 
in the hydrodynamical calculation \cite{HN04}. 
During the late stage, local thermal equilibrium is no longer 
maintained due to expansion and dilution of the matter. 
We treat this gradual transition from a locally thermalized system 
to free-streaming hadrons via a dilute interacting hadronic gas by 
employing a hadronic cascade model \cite{jam}
 below a switching 
temperature  of $T^{\mathrm{sw}}{\,=\,}169$ MeV.
A massless ideal parton gas equation of 
state (EOS) is employed in the QGP phase ($T{\,>\,}T_c = 170$ MeV) 
while a hadronic resonance gas model is used at $T{\,<\,}T_c$. When 
we use the hydrodynamic code all the way to final decoupling, we 
take into account \cite{HiranoTsuda} chemical freezeout of the hadron 
abundances at $T^{\mathrm{ch}} = 170$ MeV, separated from thermal 
freezeout of the momentum spectra at a lower decoupling temperature 
$T^{\mathrm{th}}$, as required to reproduce the experimentally 
measured yields.\cite{BMRS01} 

\begin{figure}[tbh]
\centering
\includegraphics[width=0.45\textwidth]{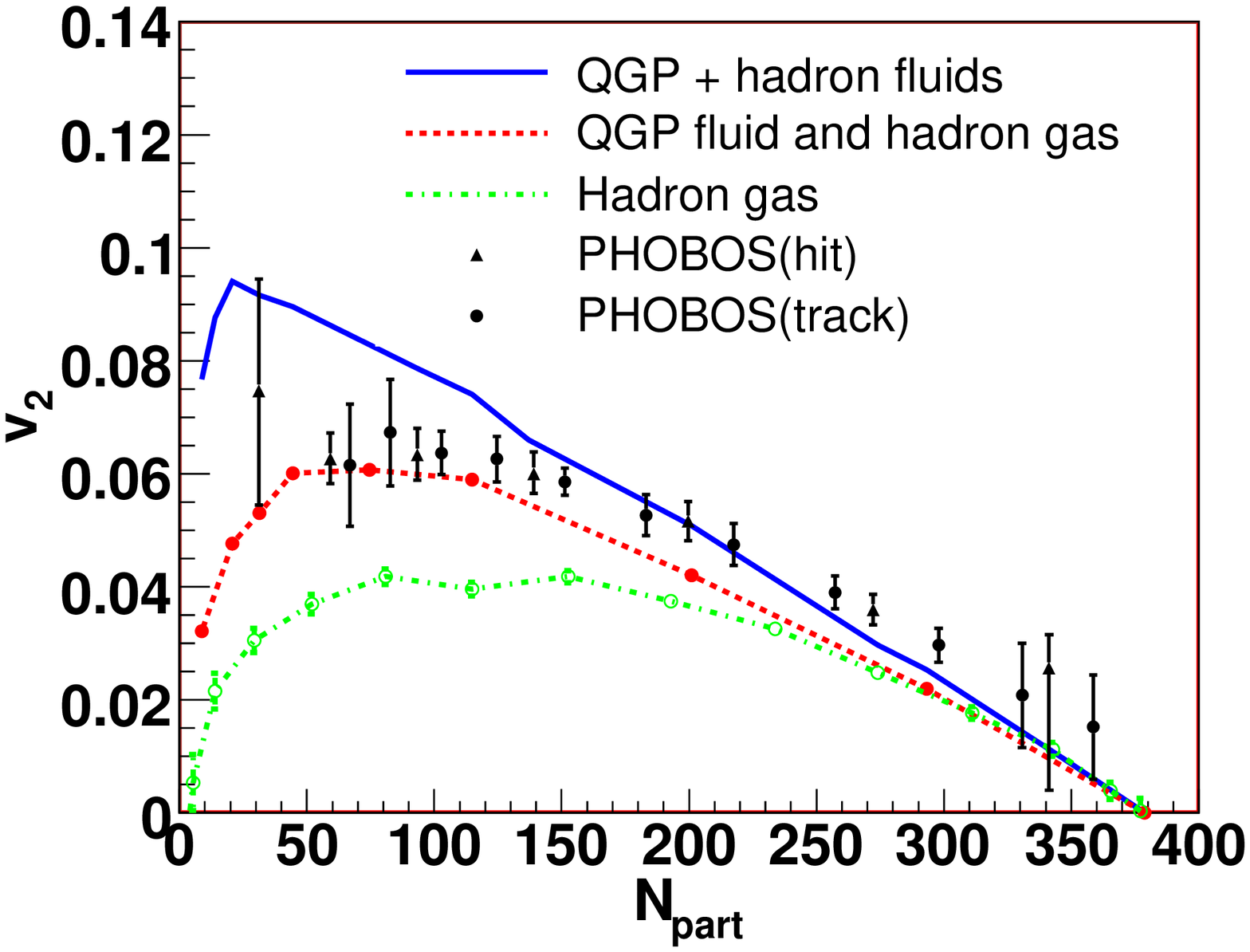}
\includegraphics[width=0.45\textwidth]{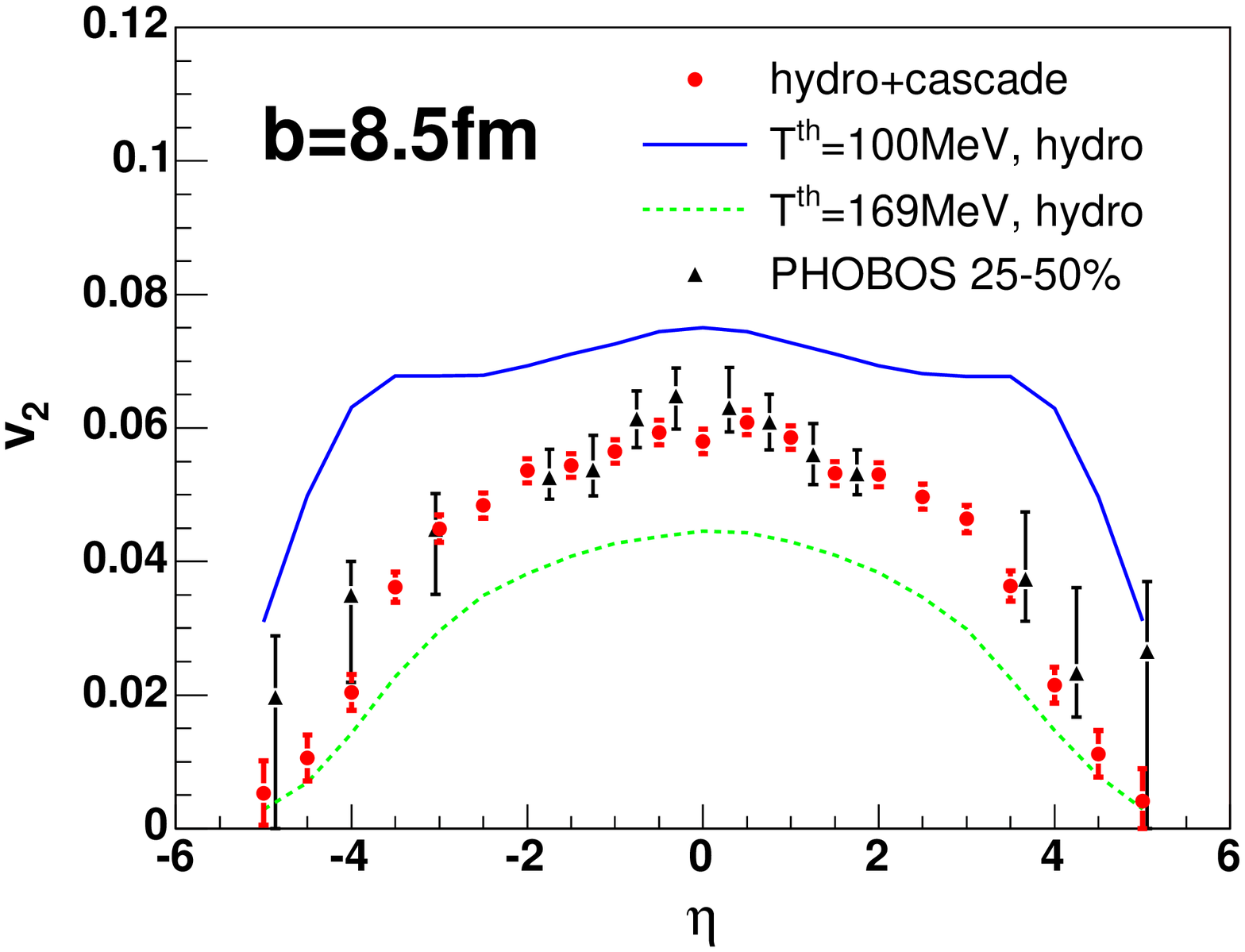}
\caption{
(Left) Centrality dependence of $v_2$. The solid (dashed) line 
results from a full ideal fluid dynamic approach (a hybrid model).
For reference, a result from a hadronic cascade model\cite{isse}
is also shown (dash-dotted line).
(Right) Pseudorapidity dependence of $v_2$. The solid (dashed) line
is the result from a full ideal hydrodynamic approach with $T^{\mathrm{th}} = 100$ MeV
($T^{\mathrm{th}} = 169$ MeV).
Filled circles are the result from the hybrid model.
All data are from the PHOBOS Collaboration. \cite{Back:2004mh}
}
\label{fig:v2npart}
\end{figure}

\section{Success of a hybrid approach}
\label{sec:success}

Initial conditions in 3D hydrodynamic simulations are put
so that centrality and pseudorapidity dependences
of charged particle yields are reproduced.
A linear combination of terms scaling with the number of participants and that of
binary collisions enables us to describe centrality dependence of
particle yields at midrapidity.
This agreement with the data still holds when the ideal fluid description 
is replaced by a more realistic hadronic cascade below $T^{\mathrm{sw}}$. 
See also Fig.~\ref{fig:cent}.
When ideal hydrodynamics is utilized all the way to kinetic freezeout,
$T^{\mathrm{th}}=100$ MeV is needed to generate
enough radial flow for reproduction of proton $p_T$ spectrum at midrapidity.  
One major advantage of the hybrid model over the ideal hydrodynamics
is that the hybrid model automatically describes freezeout processes
without any free parameters.
The hybrid model works remarkably well in reproduction of $p_T$
spectra for identified hadrons below $p_T \sim 1.5$ GeV/$c$.

The centrality dependences of $v_2$ at midrapidity ($\mid \eta \mid < 1$)
from (1) a hadronic cascade model\cite{isse} (dash-dotted),
(2) a QGP fluid with hadronic rescatterings
taken through a hadronic cascade model (dashed),
and (3) a QGP+hadron fluid with $T^{\mathrm{th}} = 100$ MeV (solid)
are compared with the PHOBOS data.\cite{Back:2004mh}
A hadronic cascade model cannot
generate enough elliptic flow
to reproduce the data.
This is observed also in other hadronic cascade calculations.\cite{hadroncascade}
Thus it is almost impossible to interpret the $v_2$ data
from a hadronic picture only.
Models based on a QGP fluid generate large elliptic flow
and gives $v_2$ values which are comparable with the data.
When a hadronic matter is also treated by ideal hydrodynamics,
$v_2$ is overpredicted in peripheral collisions.
This is improved by dissipative effects in the hadronic matter.
Note that there could exist effects of eccentricity fluctuation,\cite{fluctuation}
which is not taken into account in the current approach.
Deviation between the data and the QGP-fluid-based results
above $N_{\mathrm{part}} \sim 200$ could be attributed to these effects.

From the integrated elliptic flow data at midrapidity,
initial push from QGP pressure turns out to be important
at midrapidity.
In Fig.~\ref{fig:v2npart} (right), the pseudorapidity dependence
of $v_2$ data in 25-50\% centrality observed by PHOBOS\cite{Back:2004mh}
are compared with QGP fluid models.
Ideal hydrodynamics with $T^{\mathrm{th}} = 169$ MeV,
which is just below the transition temperature $T_{c} = 170$ MeV,
underpredicts the data in the whole pseudorapidity region.
Hadronic rescatterings after QGP fluid evolution
generate the right amount of elliptic flow and, consequently,
the triangle pattern of the data is reproduced well.
If the hadronic matter is also assumed to be described by ideal hydrodynamics
until $T^{\mathrm{th}} = 100$ MeV,
$v_2$ overshoots in forward/backward rapidity regions ($\mid \eta \mid \sim 4$).
This is simply due to the fact that, in ideal hydrodynamics, $v_2$ is approximately proportional
to the initial eccentricity which is almost independent of space-time rapidity.
So the hadronic dissipation is quite important
in forward/backward rapidity regions as well as at midrapidity
in peripheral collisions ($N_{\mathrm{part}}<100$).
From these studies, the perfect fluidity of the QGP is needed to obtain
enough amount of the integrated $v_2$, while the dissipation (or finite values
of the mean free path among hadrons) in the hadronic
matter is also important to obtain less elliptic flow coefficients
when the multiplicity is small at midrapidity 
in peripheral collisions
and/or in forward/backward rapidity regions.
This is exactly the novel picture of dynamics in
relativistic heavy ion collisions, namely, 
the nearly perfect fluid QGP core and the highly
dissipative hadronic corona, addressed in Ref.~\citen{HG05}

\begin{figure}[tbh]
\centering
\includegraphics[width=0.45\textwidth]{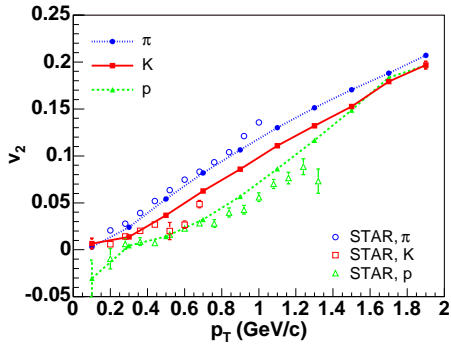}
\caption{
Transverse momentum dependence of $v_2$ for pions, kaons and protons.
Filled plots are the results from the hybrid model.
The impact parameter in the model simulation is 7.2 fm
which corresponds to 20-30\% centrality.
Data (open plots) for pions, kaons and protons 
are obtained by the STAR Collaboration.\cite{star:anisotropy}
}
\label{fig:v2pt}
\end{figure}

As a cross-check on the picture, we also study
$p_T$ dependence of $v_2$
for identified hadrons in semi-central collisions
to see whether the hybrid model works.
We correctly reproduce mass ordering behavior
of differential elliptic flow below $p_{T} \sim 1$ GeV/$c$
as shown in Fig.~\ref{fig:v2pt}.
Here experimental data are from STAR. \cite{star:anisotropy}
Although we also reproduce the data in 10-20\% and 30-40\% well (not shown),
it is hard to reproduce data in very central collisions (0-5\%)
due to a lack of initial eccentricity fluctuation
in this model. It is worth mentioning that, recently, the hybrid model
succeeds to describes differential elliptic flow data
for identified hadrons at forward rapidity observed by BRAHMS.\cite{Sanders:2007th}

\section{Challenge for a hydrodynamic approach}
\label{sec:challenge}

So far, an ideal hydrodynamic description of the QGP fluid 
with the Glauber type initial conditions
followed by an kinetic description of the hadron gas 
describes the space-time evolution of bulk matter remarkably well.
The CGC\cite{Iancu:2003xm}, whose cases are growing 
both in deep inelastic scatterings and in $d$+Au collisions recently,
is one of the relevant pictures to describe initial colliding nuclei
in high energy collisions.
In this section, novel hydrodynamic initial conditions\cite{HN04} based on the CGC
are employed for an analysis of elliptic flow.

\begin{figure}[tbh] 
\centering
\includegraphics[width=0.45\textwidth]{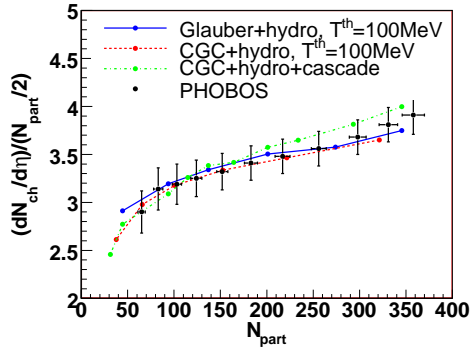} 
\caption{ 
Centrality dependence of charged particle multiplicity per number of 
participant nucleons.
The solid (dashed) line results from Glauber-type (CGC) initial conditions. 
The dash-dotted line results from our hybrid model. 
Experimental data are from PHOBOS \cite{PHOBOS}.
} 
\label{fig:cent} 
\end{figure}

We first calculate the centrality dependence of the multiplicity 
to see that the CGC indeed correctly describes the initial entropy 
production and gives proper initial conditions for the fluid dynamical 
calculations. 
Both CGC and Glauber model initial conditions, 
propagated with 
ideal fluid dynamics, reproduce the observed centrality dependence of 
the multiplicity \cite{PHOBOS}, see Fig.~\ref{fig:cent}. 
In the hydrodynamic simulations, the numbers of stable hadrons below $T^{\mathrm{ch}}$
are designed to be fixed by introducing chemical potential for each hadron.\cite{HiranoTsuda}
On the other hand, the number of charged hadrons is approximately
conserved during hadronic cascades. So 
the centrality dependence of charged particle yields is also reasonably reproduced
by the hybrid approach.

\begin{figure}[tbh] 
\begin{flushright} 
\includegraphics[width=0.45\textwidth]{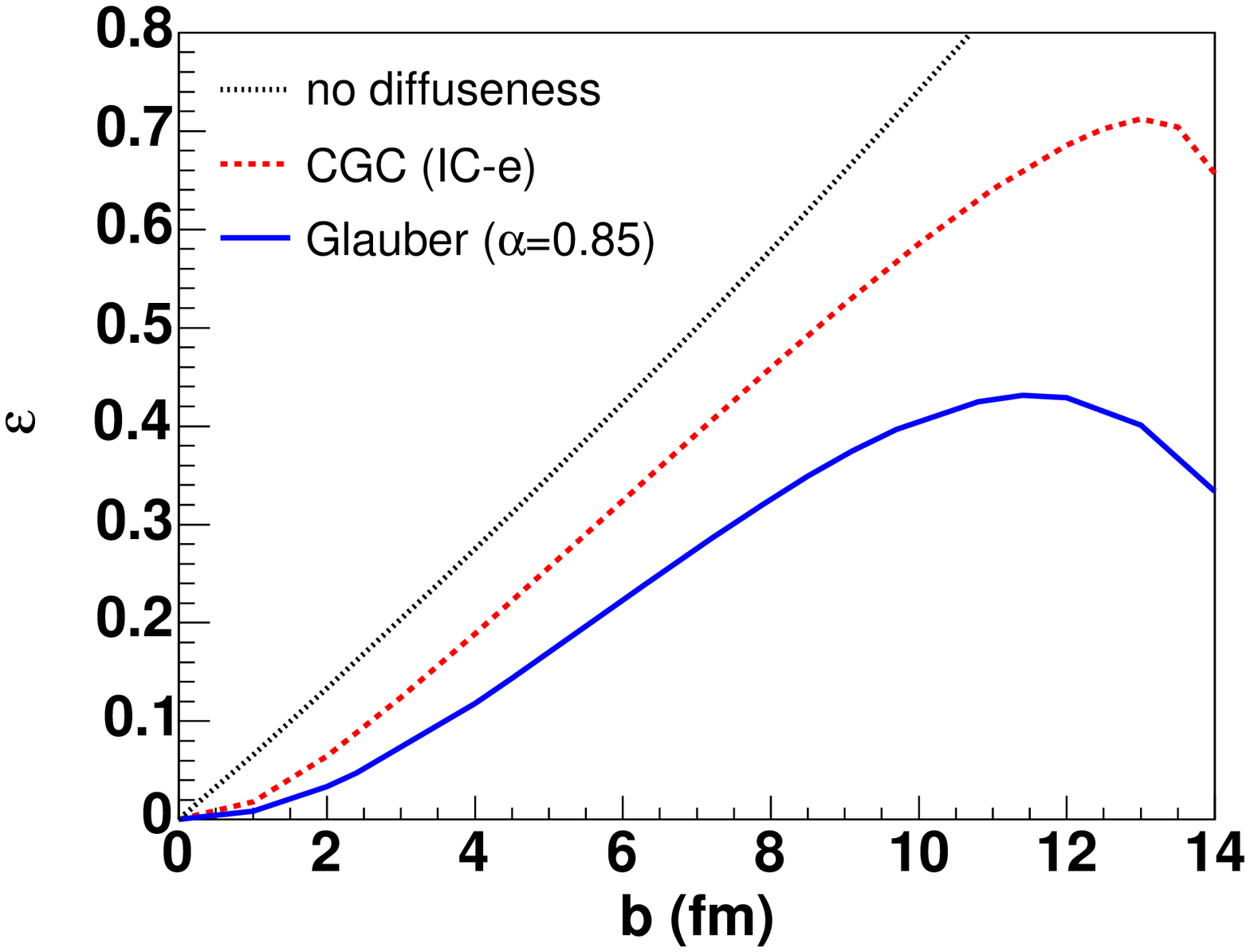} 
\includegraphics[width=0.45\textwidth]{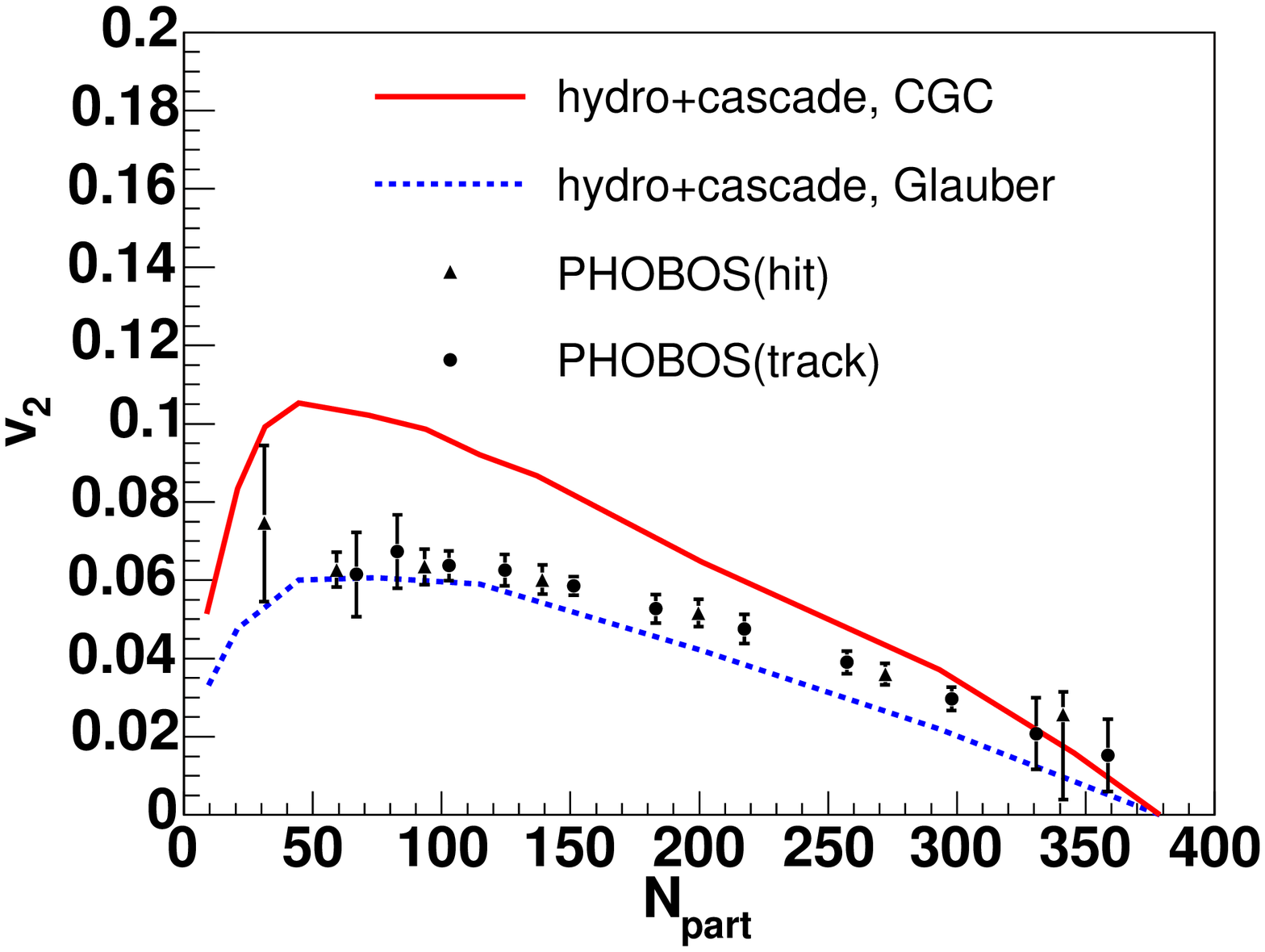} 
\end{flushright} 
\caption{ 
(Left) Impact parameter dependence of  
the eccentricity of the initial energy density distributions.
The solid (dashed) line results from Glauber-type (CGC) initial conditions. 
The dotted line assumes a box profile for the initial energy density 
distribution.  
(Right) Centrality dependence of $v_2$. The solid (dashed) line 
results from CGC (Glauber model) initial conditions followed by
ideal fluid QGP dynamics and a dissipative hadronic cascade. The data are 
from PHOBOS\cite{Back:2004mh}. 
} 
\label{fig:v2cent} 
\end{figure}

In the left panel of 
Fig.~\ref{fig:v2cent} we show the impact parameter dependence of the 
eccentricity of the initial energy density distributions at $\tau_0 = 0.6$ fm/$c$.
We neglect 
event-by-event eccentricity fluctuations although these might be 
important for very central and peripheral events \cite{fluctuation}. 
Even though both models correctly describe the centrality dependence 
of the multiplicity as shown in Fig.~\ref{fig:cent},
they exhibit a significant difference: The 
eccentricity from the CGC is 20-30\% larger than that from the Glauber 
model \cite{HHKLN,CGCecc}.
The situation does not change
even when we employ the ``universal" saturation scale \cite{Lappi:2006xc}
in calculation of gluon production.
The initial eccentricity is thus quite 
sensitive to model assumptions about the initial energy deposition 
which can be discriminated by the observation of elliptic flow. 
The centrality dependence of $v_2$ from the CGC initial conditions
followed by the QGP fluid plus the hadron gas is shown in Fig.~\ref{fig:v2cent}
(right). With Glauber model initial conditions \cite{Kolb:2000fh}, the predicted $v_2$ 
from ideal fluid dynamics overshoots the peripheral collision data 
\cite{Back:2004mh}. Hadronic dissipative effects within hadron cascade 
model reduce $v_2$ and, in the Glauber model case, are seen to be sufficient 
to explain the data (Fig.~\ref{fig:v2npart} (left)) \cite{HHKLN}. 
Initial conditions based on the CGC model, however, lead to larger elliptic 
flows which overshoot the data even after hadronic dissipation is accounted
for \cite{HHKLN}, unless one additionally assumes significant 
shear viscosity also during the early QGP stage.
Therefore precise understanding of the bulk and transport properties of QGP
from the elliptic flow data
requires a better understanding
of the initial stages in heavy ion collisions.

\section{Elliptic flow at LHC}
\label{sec:lhc}

The elliptic flow parameter plays a very important role
in understanding global aspects of dynamics in heavy ion collisions at RHIC.
It must be also important to measure elliptic flow parameter at the LHC energy
toward comprehensive understanding of the degree and mechanism of thermalization
and the bulk and transport properties of the QGP.

\begin{figure}[tbh] 
\centering
\includegraphics[width=0.45\textwidth]{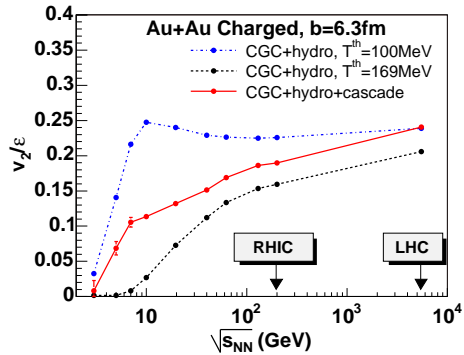} 
\caption{ 
Excitation function of $v_2/\varepsilon$ in Au+Au 
collisions at $b=6.3$\,fm.
The solid line results from CGC initial conditions 
followed an ideal QGP fluid and a dissipative hadronic cascade. The 
dashed (dash-dotted) line results from purely ideal fluid dynamics with 
thermal freezeout at $T^{\mathrm{th}}=169$\,MeV ($100$\,MeV). 
} 
\label{fig:v2epssnn} 
\end{figure}

Figure \ref{fig:v2epssnn} shows the excitation function of the 
charged particle elliptic flow $v_2$, scaled by the initial 
eccentricity $\varepsilon$, for Au+Au collisions at $b=6.3$\,fm 
impact parameter, using three different models: (i) a pure 3D ideal 
fluid approach with a typical kinetic freezeout temperature
$T^{\mathrm{th}}=100$ MeV where both QGP and hadron gas are treated as
ideal fluids (dash-dotted line); (ii) 3D ideal fluid evolution for the QGP,
with kinetic freezeout at $T^{\mathrm{th}}=169$ MeV and no hadronic 
rescattering (dashed line); and (iii) 3D ideal fluid QGP evolution
followed by hadronic rescattering below $T^\mathrm{sw}=169$\,MeV (solid line). 
Although applicability of the CGC model for SPS energies might be 
questioned, we use it here as a systematic tool for obtaining the 
energy dependence of the hydrodynamic initial conditions. By dividing 
out the initial eccentricity $\varepsilon$, we obtain an excitation
function for the scaled elliptic flow $v_2/\varepsilon$ whose shape should 
be insensitive to the facts that CGC initial conditions produce larger
eccentricities and the resulting integrated $v_2$ overshoots the data at 
RHIC and also that experiments with different collision system (Pb+Pb)
will be performed at the LHC.
Figure \ref{fig:v2epssnn} shows the well-known bump in $v_2/\varepsilon$ 
at SPS energies ($\sqrt{s_{NN}}\sim10$\,GeV) predicted by the purely 
hydrodynamic approach, as a consequence 
of the softening of the equation of state (EOS) near the quark-hadron 
phase transition region \cite{Kolb2}, and that this structure is completely 
washed out by hadronic 
dissipation \cite{Teaney}, consistent with the experimental data 
\cite{NA49,STAR}. Even at RHIC energies, hadronic dissipation still reduces 
$v_2$ by $\sim$\,20\%. The hybrid model predicts a monotonically increasing 
excitation function for $v_2/\varepsilon$ which keeps growing from RHIC 
to LHC energies \cite{Teaney}, contrary to the ideal fluid approach whose 
excitation function almost saturates above RHIC energies.\cite{Kolb2}

\section{Conclusions}

We have studied the recent elliptic flow data at RHIC by
using a hybrid model
in which an ideal hydrodynamic treatment of the QGP
is combined with a hadronic cascade model.
With the Glauber-type initial conditions,
the space-time evolution of the bulk matter
created at RHIC is well described by the hybrid model.
The agreement between the model results and the data
includes 
$v_2(N_{\mathrm{part}})$, $v_2(\eta)$,
$p_T$ spectra for identified hadron below $p_T \sim 1.5$ GeV/$c$ 
and $v_2(p_T)$ for identified hadrons at midrapidity and in the forward rapidity region.
If the Glauber type initial conditions are realized,
we can establish a picture of the nearly perfect fluid of the QGP core
and the highly dissipative hadronic corona.
However, in the case of the CGC initial conditions,
the energy density profile in the transverse plane
is more ``eccentric"
than that from the conventional Glauber model.
This in turn generates large elliptic flow,
which is not consistent with the experimental data.
Without viscous effects even in the QGP phase,
we cannot interpret the integrated elliptic flow at RHIC.
If one wants to extract informations on the properties of the QGP,
a better understanding of the initial stages is required.
We have also calculated an excitation function of elliptic flow
scaled by the initial eccentricity
and found that the function
continuously increases with collision energy up to the LHC energy
when hadronic dissipation is taken into account.

\section*{Acknowledgments}
The author would like to thank
M.~Gyulassy, U.~Heinz, D.~Kharzeev, R.~Lacey and Y.~Nara
for collaboration and fruitful discussions.
He is also much indebted to T.~Hatsuda and T.~Matsui for continuous
encouragement to the present work. 
He also thanks M.~Isse for providing him with a result from
a hadronic cascade model shown in Fig.~\ref{fig:v2npart}.
This work was partially supported by JSPS grant No.18-10104.

%


\end{document}